\documentclass[11pt,a4paper]{article}

\usepackage{graphicx}
\usepackage{gensymb}
\usepackage{amssymb}

\advance \topmargin by -\headheight
\advance \topmargin by -\headsep

\evensidemargin \oddsidemargin
\begin{document}

\begin{titlepage}

\begin{center} \Large \bf Interference of Dark Matter Solitons and Galactic Offsets

\end{center}

\vskip 0.3truein
\begin{center} 
Angel Paredes\footnote{e-mail:angel.paredes@uvigo.es}
and 
Humberto Michinel

\vspace{0.3in}

Facultade de Ciencias, \\
Universidade de Vigo, As Lagoas s/n, \\
Ourense, 32004 Spain
\vspace{0.3in}

\end{center}
\vskip 1truein

\begin{center}
\bf ABSTRACT
\end{center}

By performing numerical simulations, we discuss the collisional dynamics of stable solitary waves 
in the Schrodinger-Poisson equation. In the framework of a model in which part or all of dark
 matter is a Bose-Einstein condensate of ultralight axions, we show that these dynamics can naturally 
account for the relative displacement between dark and ordinary matter in the galactic cluster
Abell 3827,
 whose recent observation is the first empirical evidence of dark matter interactions beyond gravity. 
The essential assumption is the existence of solitonic galactic cores in the kiloparsec scale.
For this reason, we present simulations with a benchmark value of the axion mass
$m_a = 2 \times 10^{-24}$eV, which is somewhat lower than 
the one preferred for cosmological structure formation
if the field is all of dark matter ($m_a \approx 10^{-22}$eV).
 We argue that future observations might bear out or falsify this coherent wave interpretation 
 of dark matter offsets.

\vskip4.5truecm
\smallskip

\smallskip

\end{titlepage}
\setcounter{footnote}{0}

\section{Introduction}
\label{sec:intro}

The nature  of dark matter is one of the most important
open problems in fundamental physics. 
Projected experiments and astronomical observations are expected to
shed new light on this question in 
the next decade \cite{Bauer201516}.

In this context, the first evidence of dark matter (DM)  non-gravitational self-interaction 
has been recently
reported for the Abell 3827 cluster \cite{carrasco} ($z\approx 0.1$), 
where a displacement of the stars with respect to the 
maximum density of its DM halo has been observed, for some of the merging galaxies\cite{massey}.
Possible explanations  for this offset within the $\Lambda$CDM model comprise casual
allignment with other massive structures that might influence the results from
gravitational lensing, astrophysical
effects affecting the baryonic matter, tidal forces or simply wrong identification of lensed
images \cite{Williams}. Even if these causes cannot be fully excluded, meticulous observations and
simulations have shown that any such interpretation is unlikely to explain the
collected data \cite{Williams,Schaller}. This tension with collisionless dark matter models 
\cite{Schaller}
suggests the necessity
of considering other possibilities as, {\it e.g.} self-interacting dark matter, that yields
a drag force
slowing down the galactic DM distribution while leaving the standard model sector 
unaffected \cite{massey,Williams,Schaller}.
Nonetheless, 
requiring that the drag induces the offset implies a lower bound for the cross section
that is in tension with upper bounds derived from other observations,
as carefully discussed in \cite{kahlhoefer}. Thus,  the Abell 3827 cluster presents a 
challenging puzzle
that opens up questions of crucial importance to understand the nature and dynamics of DM.

In this work, we address the problem of the measured offset
 using the scalar field dark matter ($\psi$DM) model
\cite{Sin,Matos,Marsh:2015xka}, which  
 considers a Bose-Einstein condensate (BEC) of 
non-relativistic
ultra-light axions (ULAs) of mass $m_a$
subject to Newtonian gravity and that
was introduced  to solve  difficulties  of $\Lambda$CDM ({\it e.g.} 
missing satellites problem and cusp-core problem\cite{cuspcore}),
  while maintaining the successful
phenomenology of the model at cosmological scales \cite{Goodman,Hu}. 
Impressive numerical simulations \cite{schive} resolving largely different length scales have
recently given support to this expectation.
These extremely light scalar particles can arise in string theory constructions,
{\it e.g.}
\cite{Arvanitaki:2009fg} and other extensions of the standard model, {\it e.g.} 
\cite{Kim:2015yna}. Light scalars can also naturally
appear as composites
of hidden theories like the random UV field theory scenario \cite{Kiritsis:2014yqa}.

We will show that the wave-like coherent nature of BECs
severely affects the collisional dynamics of dark matter clumps, 
providing important effective forces
even in the absence of explicit local interactions between the 
elementary dark matter constituents. We then discuss the possible relevance of this
phenomenon to the puzzling observations described above.

\section{Mathematical model}
\label{sec:model}

In the condensed scalar field scenario,
the DM dynamics is governed by
a Schr\"odinger-Poisson equation\cite{PhysRev.187.1767,malomed,seidel,penrose}
for the mean-field wave-function $\psi$ of the dark matter distribution:
\begin{eqnarray}
i\, \hbar \partial_t \psi (t,{\bf x})=& -&\frac{\hbar^2}{2 m_a}\nabla^2 \psi (t,{\bf x})+
 G \, m_a^2\psi (t,{\bf x})  \int \frac{|\psi(t,{\bf x'})|^2}{|{\bf x'}-{\bf x}|}d^3 {\bf x'},
\label{SP}
\end{eqnarray}
where $|\psi|^2$ is the particle number density,
$G$ the gravitational constant and
$t$ and $\bold x$ are time and position.
For simplicity, we 
 disregard cosmological evolution of the scale factor and the contribution of baryonic
 matter to the 
gravitational field, implicitly assuming that they do not play a 
prominent role in the processes  studied below. 
Although
 a local interaction term $\lambda |\psi|^2 \psi$ can be added to 
(\ref{SP}) \cite{Goodman,PhysRevD.53.2236,guzman1}, 
we will restrict ourselves to the simplest $\lambda=0$ case
\cite{Sin,Hu, schive} that, as we show below, is enough to describe the observed
behaviour.
Notice, however, that drag forces appear in
similar mathematical models for optical systems with non-linear terms 
$\lambda \neq 0$, {\it e.g.} \cite{drag}.

Equation (\ref{SP}) can be recast in terms of adimensional quantities:
  \begin{eqnarray}
i\,\partial_t \psi (t,{\bf x})&=& - \frac{1}{2}\nabla^2 \psi (t,{\bf x}) + \Phi(t,{\bf x})
\psi (t,{\bf x})  ,\\
\nabla^2 \Phi(t,{\bf x}) &=& 4 \pi |\psi(t,{\bf x})|^2.
\label{SPadim}
\end{eqnarray}
Following \cite{schiveprl}, the adimensional unit of length, time
and mass
correspond to:
\begin{eqnarray}
\left(\frac{8\pi \hbar^2}{3 m_a^2 H_0^2 \Omega_{m0}}\right)^\frac14
&\approx& 121 \left(\frac{10^{-23}{\rm eV}}{m_a}\right)^\frac12 {\rm kpc}, \\
\left(\frac{3}{8\pi}H_0^2 \Omega_{m0}\right)^{-\frac12}&\approx& 75.5 {\rm Gyr} ,\\
\left(\frac{3}{8\pi}H_0^2 \Omega_{m0}\right)^\frac14 \frac{\hbar^\frac32}{m_a^\frac32 G}
&\approx& 7\times 10^7 M_\odot \left(\frac{10^{-23}{\rm eV}}{m_a}\right)^\frac32.
\end{eqnarray} 
We have taken $H_0=67.7$km/(s Mpc) for Hubble's constant and
$\Omega_{m0}=0.31$ for the matter fraction of energy today.

Equation (\ref{SPadim}) yields localized, radially symmetric, self-trapped robust solutions 
\begin{equation}
\psi(t,{\bf x})= \alpha e^{i\beta t} f(\sqrt{\alpha}|{\bf x}|) , \qquad
\Phi(t,{\bf x})=\alpha \varphi(\sqrt{\alpha}|{\bf x}|), 
\end{equation}
which we will loosely call solitons. $\alpha$ is an arbitrary scaling constant, the 
propagation constant is $\beta=2.454\alpha$,
the soliton mass is
$M_{sol}=\int |\psi|^2 d^3{\bf x}=3.883 \sqrt{\alpha}$ 
and its diameter (full width at half maximum) is $d_{sol}=1.380/\sqrt{\alpha}$.
$f(.)$ and $\varphi(.)$ are functions that can be computed numerically.
In terms of dimensionful quantities, the mass and size of the
solitons are related by:
\begin{equation}
M_{sol} d_{sol}\approx \frac{5.36 \hbar^2}
{m_a^2 G}\approx 4.6\times 10^{10} \left(\frac{m_ac^2}{10^{-23}{\rm eV}}\right)^{-2}
{\rm kpc} M_\odot,
\label{massdis}
\end{equation}
where $M_\odot$ is the solar mass.
In order to be reasonably self-contained,  we
give more details on these
solutions and also discuss the numerical methods
used for the computations
in the appendix (section \ref{sec:appendix}).

Finally, let us remark that
these stationary states have been independently discussed
in several physical contexts: foundations of quantum mechanics\cite{penrose,harrison},
cold trapped atoms\cite{atom1,atom2}, QCD-axions\cite{guth} and ultralight 
DM\cite{ula-soliton,Chavanis}. This often overlooked
formal coincidence indicates that studies concerning 
equation (\ref{SP}) can have deeply multidisciplinary implications. 

\section{Numerical simulations}
\label{sec:simulations}

In $\psi$DM, galactic dark matter distributions consist of a core which can be identified
with a soliton surrounded by a background also governed by Eq. (\ref{SPadim})
and evolving in time and space with uncorrelated phases\cite{schive,schiveprl,Pop}. 

In this work, we propose that the  
offset of Abell 3827 \cite{massey} can come
from  the repulsion between coherent DM  clumps
(the solitonic cores)
in phase opposition, without any extra 
local interactions.  
We show by numerical simulations that destructive
interference can provide a large effective force acting on the cores.
This repulsion  between robust wave lumps is well known in
soliton systems, from nonlinear optics\cite{solit0,solit1} to atomic 
physics\cite{matter1,matter2,matter3}, 
where the mathematical description of the phenomena is  similar to the theory 
of coherent DM waves.

In ref. \cite{massey}, observations of DM concentrations with mass of 
the order of $10^{11}M_\odot$ surrounding
stellar distributions separated by around 10 kpc were presented. 
This value of $10^{11}M_\odot$ does not correspond to a galactic mass,
but to clumps within the cluster that we will identify with solitonic cores.
Most of the mass is in the halo, which behaves incoherently and therefore 
does not feel interferential forces. We will come back to this point
in section \ref{sec:comparison}.
Taking the aforementioned values
for $M_{sol}$ and $d_{sol}$ in equation (\ref{massdis}), we find
 $m_ac^2\approx 2 \times 10^{-24}{\rm eV}$. 
 We will fix this benchmark value for the simulations below.
 In section \ref{sec:discussion}, we provide a discussion on previous observational
 constraints on $m_a$ and on their relevance to the phenomenon described here.

First, we  have analyzed the collision of two DM solitons by numerically integrating (\ref{SPadim})
with the initial condition:
\begin{eqnarray}
\psi(t=0,{\bf x}) &=&  \alpha f(\sqrt{\alpha} |{\bf x}-{\bf x_0}|) 
e^{i ({ v \cdot x})}+\alpha f(\sqrt{\alpha} |{\bf x}+{\bf x_0}|) 
e^{-i ({ v \cdot x - \Delta\phi})} 
\end{eqnarray}
where $2|{\bf x_0}|$ is the initial separation, $2v$ the initial relative velocity,
$\Delta\phi$ the relative phase and
$\alpha$ is related to normalization as described in section \ref{sec:model} 
(adimensional units).
Previous studies of this sort with $\Delta \phi=0$ can be found in \cite{guzman1,guzman2}.
We
use a split-step pseudo-spectral algorithm, known as beam propagation method 
\cite{agrawal,poon}
(see the appendix for technical details).
It is worth quoting other powerful numerical methods that have been recently developed for the 
Schr\"odinger equation 
with nonlocal terms\cite{bao,adhikari}. 

As expected, see {\it e.g.}\cite{Cavitation} for a discussion in nonlinear optics with a particular
nonlinear potential, the outcome largely depends on the relative phase and speed. 
In the case of phase opposition, 
destructive interference creates a void region between the solitons which can induce a bounce. For phase coincidence, the solitons
merge into a single matter lump (which for large initial velocities 
eventually splits again). Interference fringes appear for large velocities
 \cite{guzman1,guzman2}.

\begin{figure}[h!]
\begin{center}
\includegraphics[width=.9\textwidth]{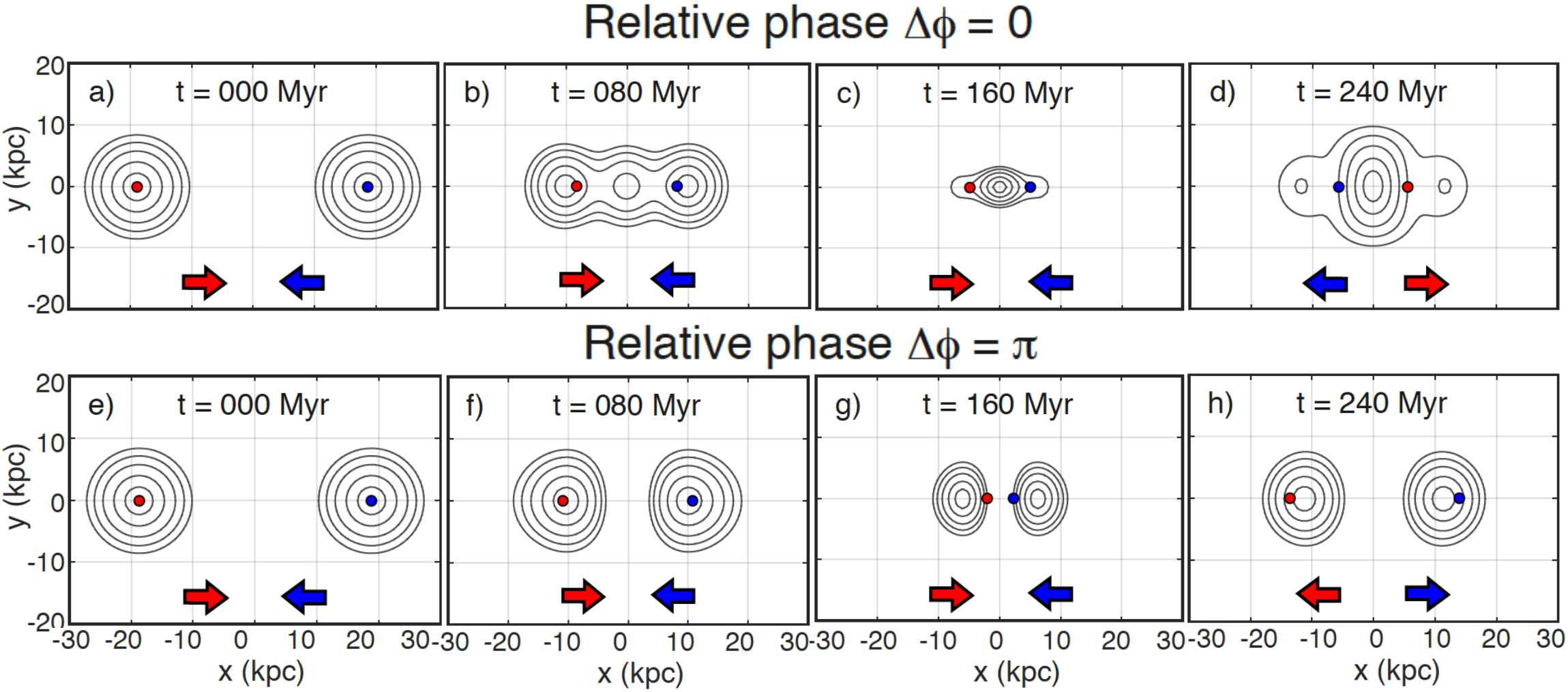}
\end{center}
\caption{ Simulation of the head-on collision of two  DM solitons with
$M_{sol} = 10^{11} M_{\odot}$,
 centers initially separated by 40 kpc and 
  relative velocity 200 km/s. The contours show the projected DM
mass density integrated over $z$. 
 Dots are the point particles representing the center of gravity of ordinary 
matter in each lump and arrows indicate the direction of their velocity. 
Panels a)-d) show different instants of a simulation in which the  solitons 
are launched in phase coincidence and  merge. The sequence e)-h)
 corresponds to phase opposition and the DM clouds bounce back.}
\label{fig1}
\end{figure}

\begin{figure}[h!]
\begin{center}
\includegraphics[width=.5\textwidth]{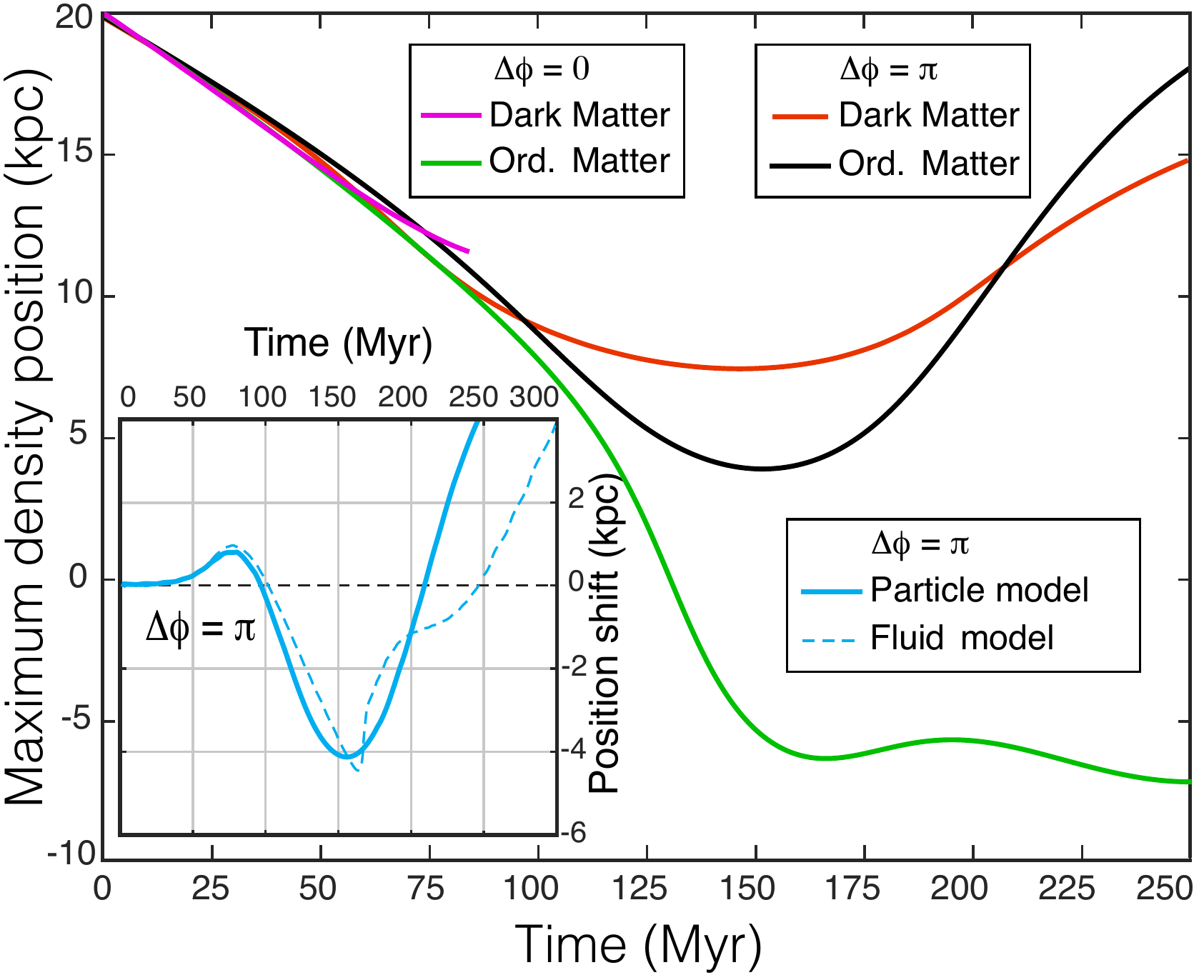}
\end{center}
\caption{Time evolution of the position of the maximum density of one of the DM 
solitons and 
its corresponding test particle representing the stars
for the simulations of figure \ref{fig1}. For the particular cases studied $\Delta \phi=0,\pi$,
the dynamics for the other soliton is symmetric. 
For the collision in phase opposition, corresponding to the right 
column of figure \ref{fig1}, offsets between the maxima of DM and
ordinary matter are generated dynamically.
The time evolution of the relative displacement is shown in the inset. 
In the case of phase coincidence ($\Delta\phi=0$) no 
significant
offset is observed. 
For $t > 85$Myr, the solitons have merged and the
DM maximum lies at the center and thus, the pink line is cut at this point. 
}
\label{fig2}
\end{figure}

We must underline that in this work, for the first time to our knowledge, 
the effect  of coherent DM waves 
on luminous matter has been calculated, by adding to our simulations test particles following classical trajectories in the gravitational field generated 
by the DM wave. These particles, initially located at the soliton centers, are a toy representation 
of the stars and can be shifted from the DM density peaks in
a collision, as we show in figure 
\ref{fig1}. In figure 
\ref{fig2}, 
we plot the comparison between the trajectories of the point particle and the DM 
projected mass maximum. In order to check the limitations of this 
particle model, we have made use of the
well known fact that 
 Schr\"odinger equation can be cast into a hydrodynamic form through the Madelung transformation \cite{fluid_nlse}. This allows us to develop a fluid toy model in which luminous matter is
 described as a spatially extended cloud (see the appendix). As it can be seen in 
 the inset of figure \ref{fig2}, both models  display a good qualitative agreement.
 
\begin{figure}[h!]
\begin{center}
\includegraphics[width=.9\textwidth]{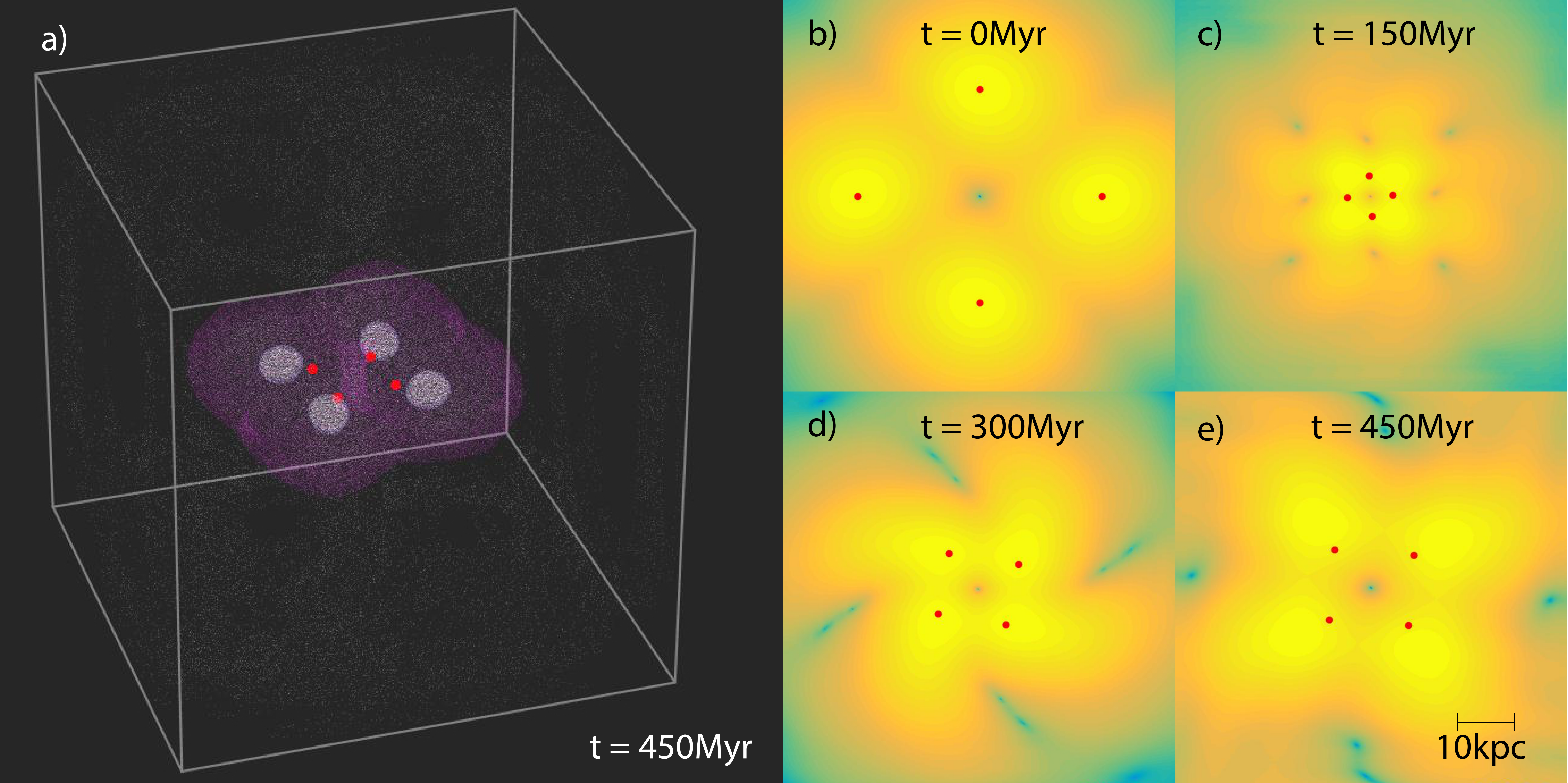}
\end{center}
\caption{Simulation showing an example of ordinary matter vs. DM offset in a vortex-like
configuration.  In a), 
we show a three dimensional representation of the system at time $t=450$Myr. 
The light and purple blobs are density iso-surfaces at 64$\%$ and 4$\%$ of the initial
maximum density. Plots b)-e) are density color maps in logarithmic scale for the $z=0$ plane
at the times indicated in each picture. In all plots, the red dots are the point particles representing ordinary
matter. The offset can be clearly appreciated at  $t=150$Myr and $t=450$Myr.
}
\label{fig3}
\end{figure}

Even if the collision in phase opposition is the simplest case, 
luminous vs. DM shifts can happen in more general situations. 
Figure 
\ref{fig3} 
shows an example with four galaxies. 
Initial conditions are four solitons of mass $M_{sol}=0.72 M_\odot$ each, located
at the vertices of a square of diagonal 40 kpc and initial velocities of 100 km/s towards the
center, with phases 0, $\pi/2$, $\pi$ and $3\pi/2$, respectively. When the solitons approach 
each other, their phase gradients induce a rotation of the DM cloud, with ordinary matter 
lagging behind. It is worth mentioning that stationary rotating solutions of
Eq. (\ref{SP}) have been discussed in \cite{shapiro,guzman3}.

Interference also plays an important role  in asymmetric  collisions if the phase difference between the
lumps remains a well-defined quantity during the process. As an estimation, take $|\beta_1 - \beta_2| \Delta t_{col} \lesssim 1$
where $\beta_1$ and $\beta_2$ are the propagation constants of each soliton  and
$\Delta t_{col}$ is the duration of the collision. 
In terms of the soliton mass: 
\begin{equation}
\beta({\rm Myr}^{-1})=4.33 \left(\frac{M_{sol}}{10^{11}M_\odot}\right)^2
\left(\frac{m_ac^2}{10^{-23}{\rm eV}}\right)^3.
\label{betadim}
\end{equation}
 This yields:
\begin{equation}
 \Delta t_{col}({\rm Myr})\left(\frac{|M_{sol,1}|^2-|M_{sol,2}|^2}{(10^{11}M_\odot)^2}\right)\left(\frac{m_ac^2}{10^{-23}{\rm eV}}\right)^3 \lesssim \frac{1}{4.33}. 
 \label{condition}
\end{equation}
In section \ref{sec:comparison}, a collision involving four solitons will be considered.
It is easy to check that the condition (\ref{condition}) holds in that case.

The relative phase between two solitons is a decisive factor for their
collisional dynamics, and, in the $\psi$DM model, it is important for galactic mergers.
This phase is ultimately determined by initial conditions at galactic formation. Moreover, the
relative phase for any pair solitons changes in time   since the propagation
constant $\beta$ depends on the mass, Eq. (\ref{betadim}). 
In theory, given precise initial conditions, the cosmological evolution of the axion field can be computed
deterministically in the semiclassical description of Eq. (\ref{SP}) \cite{schive}.  
In practice, uncertainties in the initial conditions and the evolution imply that the relative
phase for a particular collision can actually be considered as random.

\section{Comparison with observations}
\label{sec:comparison}
\begin{figure}[htb]
\begin{center}
\includegraphics[width=.5\columnwidth]{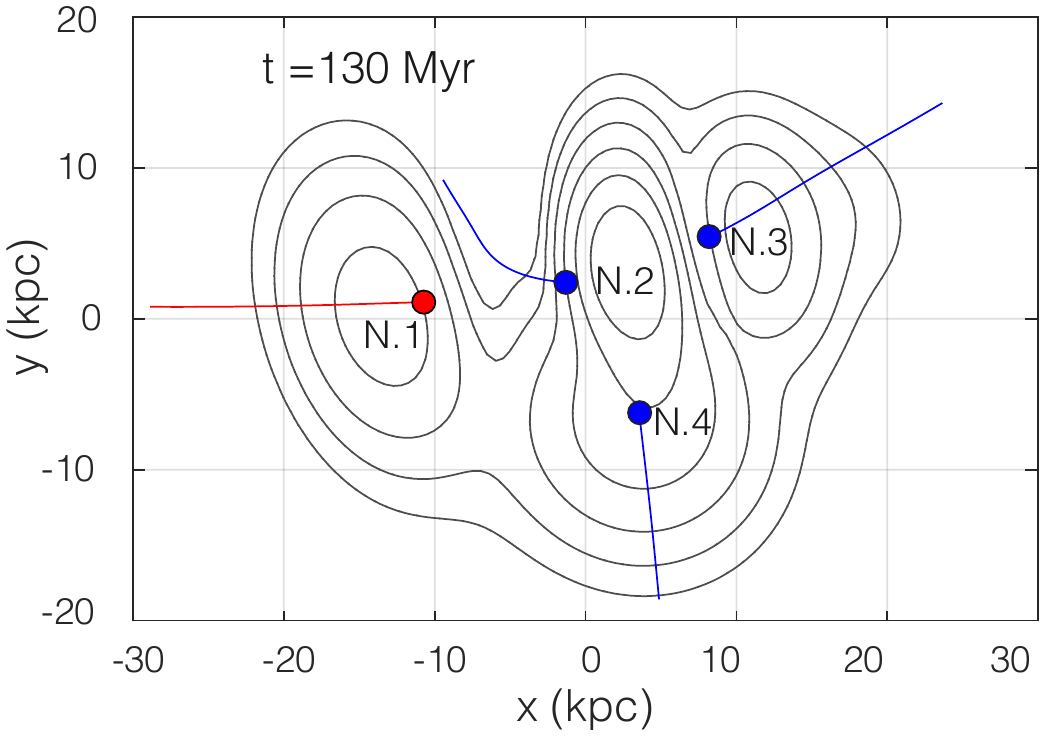}
\end{center}
\caption{A numerical simulation generating 
an offset similar to the Abell 3827 cluster.
The contour plot shows the projected mass density at a given time. 
Dots (numbered from N.1 to N.4) 
are the point particles representing the center of gravity of ordinary matter 
in each lump. Color lines indicate their trajectories 
 from $t=0$ to $t=130$Myr. 
The qualitative agreement with observations reported in  ref.\cite{massey},
including the offset of N.1,
is remarkable.}
\label{fig4}
\end{figure}

We now show that, starting with separate solitons, the wave dynamics of equation (\ref{SP}) can
generate the gross features of the Abell 3827 cluster\cite{massey}: 
there are two DM blobs, one
comprising galaxy N.1, for which dark matter and stars are separated;
and the other one comprising galaxies N.2-N.4. 
In the present scenario, the natural interpretation is that N.1 is in phase
opposition to N.2 whereas N.3 and N.4 are in phase. Figure 
\ref{fig4} 
 shows the result
of a simulation. Initially, four separate solitons with masses 0.72, 0.95, 1.28 and 1.1 times
$10^{11}M_\odot$ are considered.
Solitons 1 and 3 are heading soliton 2 with
relative velocities of 220 and 180 km/s. 
Soliton 4 has an initial velocity of 900 km/s in the transverse direction, in order to
agree with the redshift measured in\cite{massey}. 
After evolution, we find offsets similar to those displayed in \cite{massey}.
Matching these
qualitative features as in figure 
\ref{fig4} obviously requires an appropriate choice of initial conditions
but we remark that no special fine tuning is needed.

Obviously, the real  conditions are far more complicated. Apart from the coherent
solitonic core,  DM of field galaxies includes a non-coherent halo with an approximate 
Navarro-Frenk-White profile, 
see\cite{schiveprl} for a detailed discussion. When the cluster is formed, most of its matter will be
in an incoherent state with, at most, coherent lumps around the initial galactic cores (namely,
around the stellar distributions). However, the presence of a large incoherent background
does not necessarily change the qualitative features of the dynamics. Clearly, the effect of
soliton-cluster halo
interferences averages out to zero and can be neglected. Moreover,
since the background density varies
only mildly within the cluster, the gravitational forces it generates will not be dominant.
A natural concern is whether incoherent matter might be attracted by the larger densities
at the solitons, leading to smaller and more massive lumps. This is avoided 
if the kinetic energy of the incoherent wave is enough to impede
its absorption, as  in\cite{schiveprl} for single galaxies.
In fact, we have checked by numerical simulation that an incoherent background does not severely affect
the process of figure \ref{fig4}.

\section{Discussion on the axion mass}
\label{sec:discussion}

In this section, we briefly review the observational constraints on the axion mass
$m_a$ (see {\it e.g.} \cite{Pop} and references therein)
and their relation to our dark wave interpretation of the offsets.
The essential hypothesis for our modeling is the 
existence of kiloparsec scale coherent cores.
We remark that the dark matter density distributions for
$R  \lesssim 5-6$kpc of galaxies like the Milky Way are 
subject to large uncertainties
\cite{Pato:2015tja},
 \cite{Calore:2015oya} and therefore the assumption is neither confirmed nor
excluded by direct inference of the galactic profiles.

It is  natural
to assume that we are in a scenario in which the cusp-core problem is  solved 
solely by $\psi$DM.
By studying the Fornax dwarf galaxy in this context,
the authors of ref.
\cite{schive} found a 
best fit of $m_ac^2=(8.1^{+1.6}_{-1.7})\times 10^{-23}$eV. 
A related analysis of Fornax and Sculptor in ref. 
\cite{Pop} yielded   a one-sided constraint 
$m_ac^2<1.1\times 10^{-22}$eV. See also \cite{Marsh:2015xka}
and references therein.

On the other hand,
a  lower bound $m_a c^2 > 10^{-24}$eV comes 
from requiring that $\psi$DM is indistinguishable from $\Lambda$CDM  
for the probes studied in\cite{Grin}. This is a conservative 
lower bound, derived only from linear constraints on the cosmic 
microwave background. More stringent but also more model dependent
lower bounds were derived 
from nonlinear probes
in \cite{marsh1,marsh2,schive2015},
see also \cite{Marsh:2015xka}, \cite{Amendola:2005ad} and 
references therein. 
Let us quote the result of \cite{marsh2},
where it is found that data from the Hubble Ultra-Deep Field exclude
axions with $m_a \lesssim 10^{-23}$eV contributing more than half of DM.
Signals from pulsars might soon give new information on the existence of
ultralight axions and their mass \cite{1475-7516-2014-02-019}.

The benchmark value of $m_a=2 \times 10^{-24}$eV 
that we have fixed in the simulations
shown in the sections \ref{sec:simulations} and \ref{sec:comparison} 
comes from requiring solitonic cores
with radius of the order
of few kiloparsecs for masses of the order of $10^{11}M_\odot$.  
We allow ourselves to use this value of $m_a$ since it complies with
the conservative lower bound of \cite{Grin}. 
However, we envisage two possibilities in which
these large cores could be present for larger values of $m_a$:

First, ULAs could be just a fraction of dark matter, relaxing  to some extent the
aforementioned stringent
mass constraints, see {\it e.g.} \cite{marsh2}. Moreover, the total mass 
constituting each solitonic
core would be smaller (for a fixed total dark matter mass), leading to larger
radii by virtue of Eq. (\ref{massdis}). 
The mechanism introduced in this paper can
only cause a displacement of the ULA fraction of dark matter
from the stars, but that can anyway render an offset for the 
DM center of mass.

Second, if a repulsive term $\lambda |\psi|^2 \psi$, as first introduced
in \cite{Goodman,PhysRevD.53.2236}, is added to Eq. (\ref{SP}), the soliton
radius would be larger than the one given in Eq. (\ref{massdis}),
see \cite{atom2} for a detailed discussion.

If future observations and/or analysis  indicate that $\psi$DM can only be realized with 
subkiloparsec cores for Milky Way-class galaxies,
it would then be unlikely that soliton interactions
 play any role for providing relevant offsets
within clusters like Abell 3827.
Nevertheless, the analysis in this work would still play a role for the 
interaction between the solitonic cores. 
Understanding whether it might yield observational
consequences is left for the future.

\section{Conclusions}
\label{sec:conclusions}
We have discussed the phenomenon of soliton interactions based on 
wave interference, which is relevant for any model of BEC
dark matter relying on a Schr\"odinger-Poisson equation
\cite{Sin,Matos,Marsh:2015xka,guth,PhysRevLett.103.111301}.
Large effective forces can be induced during
collisions, in  analogy with well known experiments in laboratory BECs and
nonlinear optical systems. 

 If an ultralight scalar represents a significant fraction of DM, 
 it is plausible that interference between dark waves
can have observational consequences for galactic mergers and, in particular, it can
explain the consequential results of\cite{massey}. 
The simplest setting for generating offsets is that of head-on collisions
in phase opposition
(see figures \ref{fig1} and \ref{fig2}), but we stress that they appear in more
general situations (figures \ref{fig3} and \ref{fig4}).
The  paramount hypothesis is the existence of  coherent cores with
radii of the order of few kiloparsecs.
There are important  qualitative differences with other models of DM:
 the force acting on DM is between the solitonic cores
 and not between a galaxy and the cluster halo. 
Moreover, the outcome depends on the 
 value of the relative phase at the moment of the collision, which in a realistic 
 situation could be taken as random. 
These two points can be
tested if other similar mergers are observed with the level of detail achieved by 
\cite{massey} and can potentially reconcile the offset in \cite{massey} with
the lack thereof in other systems \cite{bullet,massey2}, which are in tension
in models with particle-like interactions \cite{kahlhoefer}.
It is worth emphasizing that the dark matter shift
 discussed here is different from the one observed in
other systems like the Bullet cluster \cite{bullet}, where the offset is between dark matter and
gas, not stars, it is observed after the halos have traversed each other and is perfectly
consistent with collisionless dark matter. 
A natural question is whether interference effects
could then spoil the standard description of those systems. That is unlikely because the
effective forces only act on the solitonic cores and are therefore confined to the kiloparsec 
scale or less. We expect the corrections to average out to zero in larger collisions like
the Bullet cluster, with a size of a few megaparsecs. Direct numerical confirmation of
this assertion is left for future work.

In the present work, we have considered an extremely simplified description of galactic
dynamics which is enough to understand the gross features that can be expected
from a wavelike behavior of DM. 
It would of course be desirable to incorporate these features in more
detailed simulations as, for instance,
 those reported in \cite{Calore:2015oya} or \cite{massey2}.

Thus, we expect that
through future theoretical progress and 
astrophysical observations, the scientific community will be able 
to discern the present scenario from models with explicit DM self-interactions or other
logical possibilities. In a broader perspective, it is worth emphasizing that continuously improving
observational evidence increasingly calls for precise descriptions of nonlinear phenomena.
For instance \cite{Pontzen:2015eoh} has studied the consequences of nonlinear evolution in the formation
of cosmic voids. It is of great interest to understand whether alternatives to $\Lambda$CDM
lead to differences that might  be experimentally tested. Whether the discussion of the present
contribution or the $\psi$DM model in general may have implications in this respect is a 
compelling question for the future.

Finally, it is worth pointing out that,
due to the formal coincidence of the governing equations,
refined control of trapped atoms\cite{atom1} and
optical media\cite{solit2} introducing
gravity-like interactions might  allow for laboratory analogue simulators
of galactic-scale phenomena.

\section{Appendix: Technical details on numerical methods}
\label{sec:appendix}

In this appendix we describe a number of technical issues related
to the numerical treatment of the Schr\"odinger-Poisson equation. We will use
the  form of the equation in terms of dimensionless quantities (\ref{SPadim}).

\subsection*{ Stationary solution and initial conditions} 

\begin{figure}[ht!]
\begin{center}
\includegraphics[width=.5\textwidth]{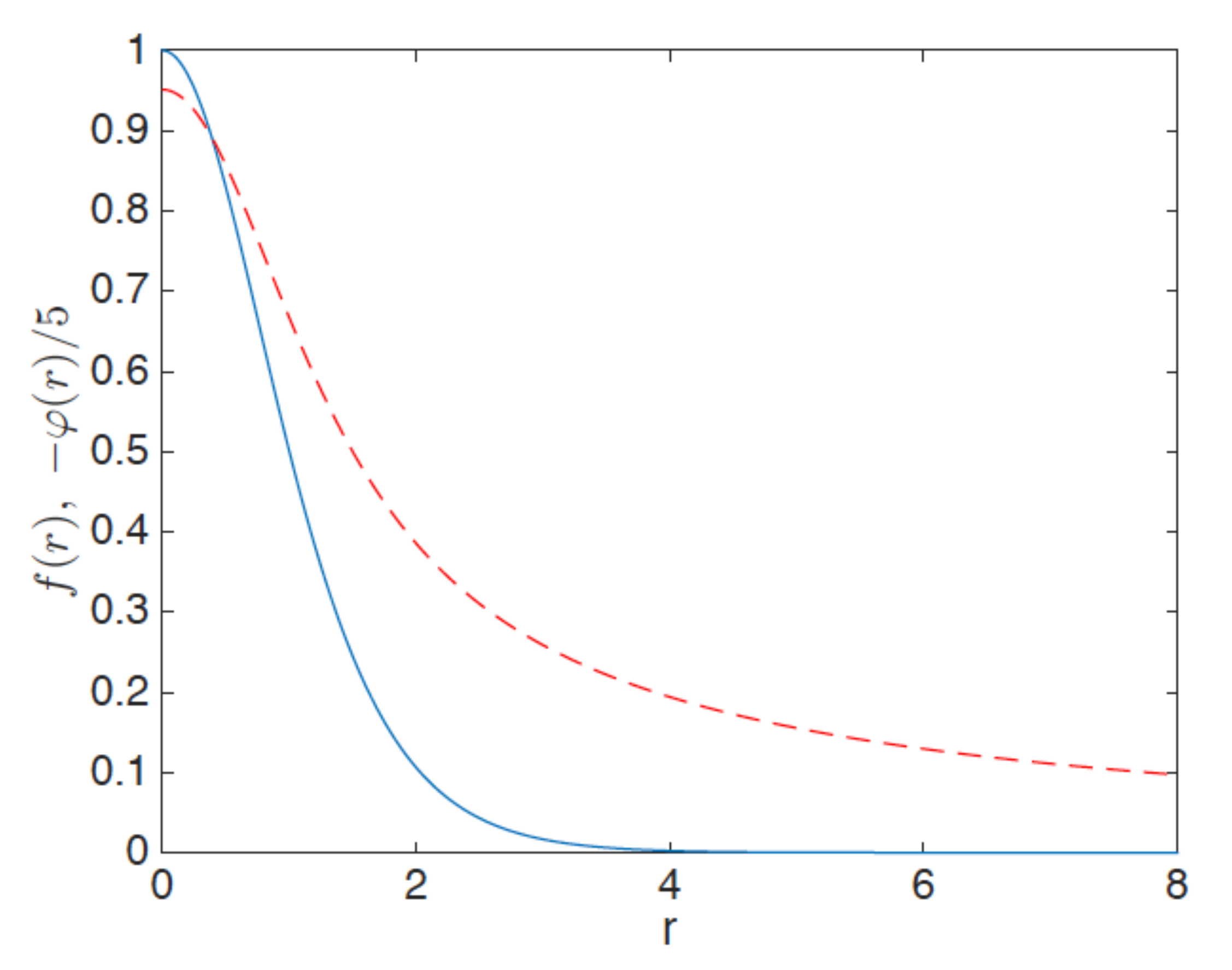}
\end{center}
\caption{The functions $f(r)$ (solid blue) and $-\varphi(r)/5$ (dashed red)
as defined in the text. 
}
\label{fig_soliton}
\end{figure}

Consider an ansatz with radial symmetry $\psi(t,{\bf x})= e^{i\beta t} f(r)$,
$\Phi(t,{\bf x})=\varphi(r)$ where
we have defined $r = |{\bf x}|$.
Equation (\ref{SPadim}) is reduced to:
\begin{eqnarray}
0 &=& -\frac12 \frac{d^2f(r)}{dr^2}- \frac{1}{r}\frac{df(r)}{dr}+ \varphi(r) f(r) +\beta f(r)\nonumber\\
0&=&\frac{d^2 \varphi(r)}{dr^2}+\frac{2}{r}\frac{d\varphi(r)}{dr}- 4 \pi f(r)^2
\end{eqnarray}
Moreover, $\beta$ can be reabsorbed as $\tilde \varphi(r) = \varphi(r) + \beta$.
In order to find the soliton solution, we fix $f(r=0)=1$ and perform a standard shooting
method by varying $\tilde \varphi(r=0)$. 
Regularity at $r=0$ ensures that there are no more free parameters and
$\tilde \varphi(r=0)$ is fixed by requiring that $f(r)$ vanishes as $r\to \infty$.
Since $\lim_{r\to \infty}\varphi(r) = 0$, the value of $\beta$
is read from the large $r$ behavior of $\tilde \varphi(r)$. 
We find that $\beta=2.454$, $M_{sol}=4\pi\int_0^\infty r^2 f(r)^2 dr = 3.883$
and the full width at half maximum of the density is  fwhm$\equiv d_{sol}$=1.380. 
The solution is depicted in figure \ref{fig_soliton}. Notice that
$\varphi(r)$ has been rescaled in order to refer both functions to the same axis
($\varphi(r=0)\approx -4.76$).

Due to the symmetry
of equation (\ref{SPadim}), a whole family of solutions can be found by scaling.
If we take $|\psi|(r=0)=\alpha$ for any $\alpha>0$, then there is a solitary wave solution
with $M_{sol}=3.883 \sqrt{\alpha}$, $\beta=2.454\alpha$ and fwhm$=1.380/\sqrt{\alpha}$.

Once a stationary solution is found and the functions $f(r)$, $\varphi(r)$ are explicitly known,
Galilean invariance of the Schr\"odinger-Poisson equation allows us to write down the
general
boosted
solution representing a soliton moving with constant velocity ${\bf v}$:
\begin{eqnarray}
\psi(t,{\bf x})&=&\alpha f(\sqrt{\alpha}|{\bf x}-{\bf v}t|)
e^{i (\alpha \beta t +{\bf v}\cdot{\bf x}-\frac12 |{\bf v}|^2t)} \nonumber\\
\Phi(t,{\bf x})&=&\alpha \varphi(\sqrt{\alpha}|{\bf x}-{\bf v}t|)
\label{moving_soliton}
\end{eqnarray}

This  solution is used to define the initial conditions for the simulations, in which
we consider initially separated solitons.
For instance, the soliton collision of figure 1 has initial condition:
\begin{equation}
\psi(t=0,{\bf x})= \alpha f(\sqrt{\alpha}\sqrt{(x+x_0)^2+y^2+z^2})
e^{i v x} +
\alpha f(\sqrt{\alpha}\sqrt{(x-x_0)^2+y^2+z^2})
e^{-i v x + i \Delta \phi}
\end{equation}

 where ($-x_0$, $v$), ($x_0$,$-v$) are the (adimensional) initial position and velocity of
 each soliton, $\Delta \phi$ their relative phase and $3.883\sqrt{\alpha}$ the adimensional
 soliton mass. 
 The test particles modeling the stars are initially placed at the center of the solitons
  with the same initial velocity.
 It is worth noticing that, due to Galilean invariance of equation
 (\ref{SPadim}), the
 same process can be thought of as, for instance, one soliton initially placed at $-2x_0$
 moving with velocity $2v$ toward a static soliton with center at $x=0$. However, 
 it is necessary to
 change  the initial phase accordingly, namely:
 \begin{equation}
\psi(t=0,{\bf x})= \alpha f(\sqrt{\alpha}\sqrt{(x+2x_0)^2+y^2+z^2})
e^{ 2i v x} + 
\alpha f(\sqrt{\alpha}\sqrt{x^2+y^2+z^2})
e^{i( \Delta \phi- 2 v x_0)}
\end{equation}
Similarly, the initial condition used for the simulation leading to figure 4
is, explicitly:
\begin{equation}
\psi(t=0,{\bf x}) = \sum_{n=1}^4 \alpha_n f(\sqrt{\alpha_n} |{\bf x}-{\bf x_n}|) 
e^{i ({\bf v_n \cdot x_n} + \phi_n)}
\end{equation}
with $\alpha_1=544$, ${\bf x_1}=(-0.175,0.040,-0.060)$,
${\bf v_1}=(67.4,-15.4,69.2)$, $\phi_1=3.17$, $\alpha_2=947$, ${\bf x_2}=(-0.027,0.040,-0.060)$,
${\bf v_2}=(5.04,-15.4,69.2)$, $\phi_2=0$, $\alpha_3=1720$, ${\bf x_3}=(-0.120,0.040,-0.060)$,
${\bf v_3}=(-46.0,-15.4,69.2)$, $\phi_3=0.796$, $\alpha_4=1270$, ${\bf x_4}=(-0.002,-0.108,0.161)$,
${\bf v_4}=(5.04,41.3,-186)$, $\phi_4=0$. The plot displayed in figure 4
 was rotated in the x-y 
plane
for a better visualization.

\subsection*{ Dynamical evolution}

In order to compute temporal evolution in (\ref{SPadim}), 
we have used a split-step Fourier method, 
widely used in the integration of the nonlinear Schr\"odinger equation because of
its stability and precision, see for instance references \cite{agrawal} and \cite{poon}. 
Schematically $\psi(t+dt,{\bf x}) = F^{-1}[e^{-i {\bf k}^2 dt/2}
F[e^{i \Phi dt} \psi(t,{\bf x})]]$ where $F$ is the three-dimensional Fourier transform.
The preservation of the norm $\int |\psi|^2 d^3{\bf x}$ is automatic.
At each step, we need to compute $\Phi(t,{\bf x})$ from Poisson equation.
We do so using
a finite difference scheme
in a $N_x \times N_y \times N_z$ spatial grid.
The discrete Poisson equation can be written as a linear problem
$A \cdot B = C$ where $A$ is the 
corresponding heptadiagonal $N_xN_yN_z \times N_xN_yN_z$ sparse matrix,
$B$ is a $N_xN_yN_z \times 1$ matrix with the value of the function
$\Phi$ at each point of the grid and $C$ includes the source and the boundary terms.
This algebraic problem can be efficiently solved by an iterative symmlq method.
We introduce boundary conditions as if all the mass were concentrated at the center of the
computational grid (we compute in a reference frame in which the center of mass coincides
with the
center of the grid 
and the total momentum vanishes). The error introduced by these boundary conditions
becomes negligible as the size of the computational box becomes much larger than the
size of the region of interest. 
As an additional cross-check, we have also used a second method for solving Poisson
equation, namely that of directly using Fourier transformation to deal with
the laplacian. That 
implies periodic boundary conditions for $\Phi$, which also approach the physical 
boundary conditions as the box is made larger.
In order to compute the classical trajectories for the
point particles representing the standard model matter, we use Heun's algorithm
(a second order Runge-Kutta scheme), making use of the gravitational potential computed
at each time step.

\begin{figure}[ht!]
\begin{center}
\includegraphics[width=.6\columnwidth]{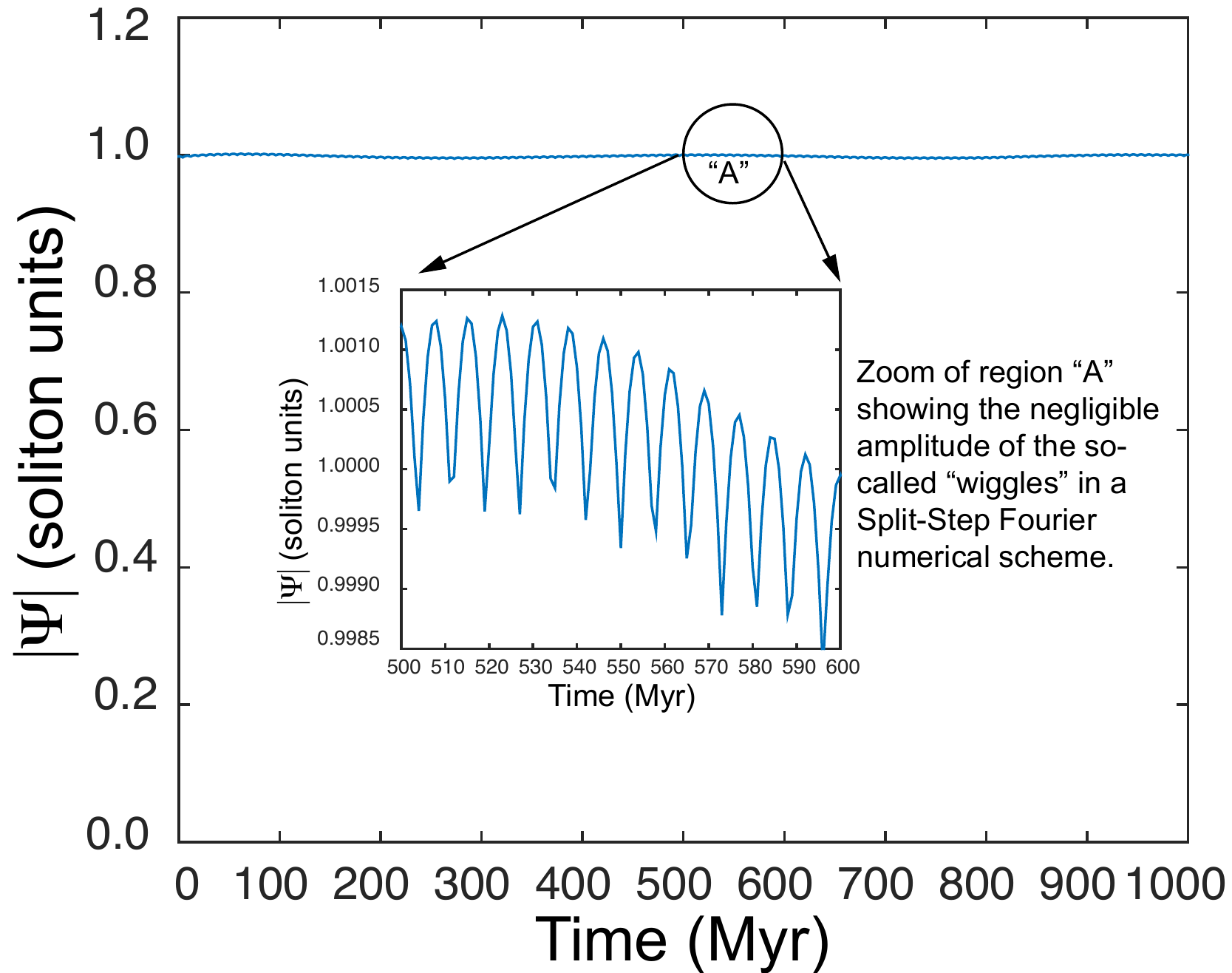}
\end{center}
\caption{Simulation showing that the effect of spurious "wiggles" due to the initial velocity of the galaxies is negligible in a split-step Fourier numerical scheme.
}
\label{suppl_fig3}
\end{figure}

In order to test the precision of the algorithm, we first track the solution corresponding
to a soliton moving with constant velocity, see figure \ref{suppl_fig3}, where the evolution of the
modulus of the wave function at the center of the soliton is plotted. 
Numerical errors introduce two types of oscillations around the theoretical constant
value. The short period oscillation is due to the velocity whereas a fluctuation with a
a longer period appears also for ${\bf v}=0$. In any case, the figure shows that the
method is stable and that fluctuations can be kept small.

\begin{figure}
\begin{center}
\includegraphics[width=.49\columnwidth]{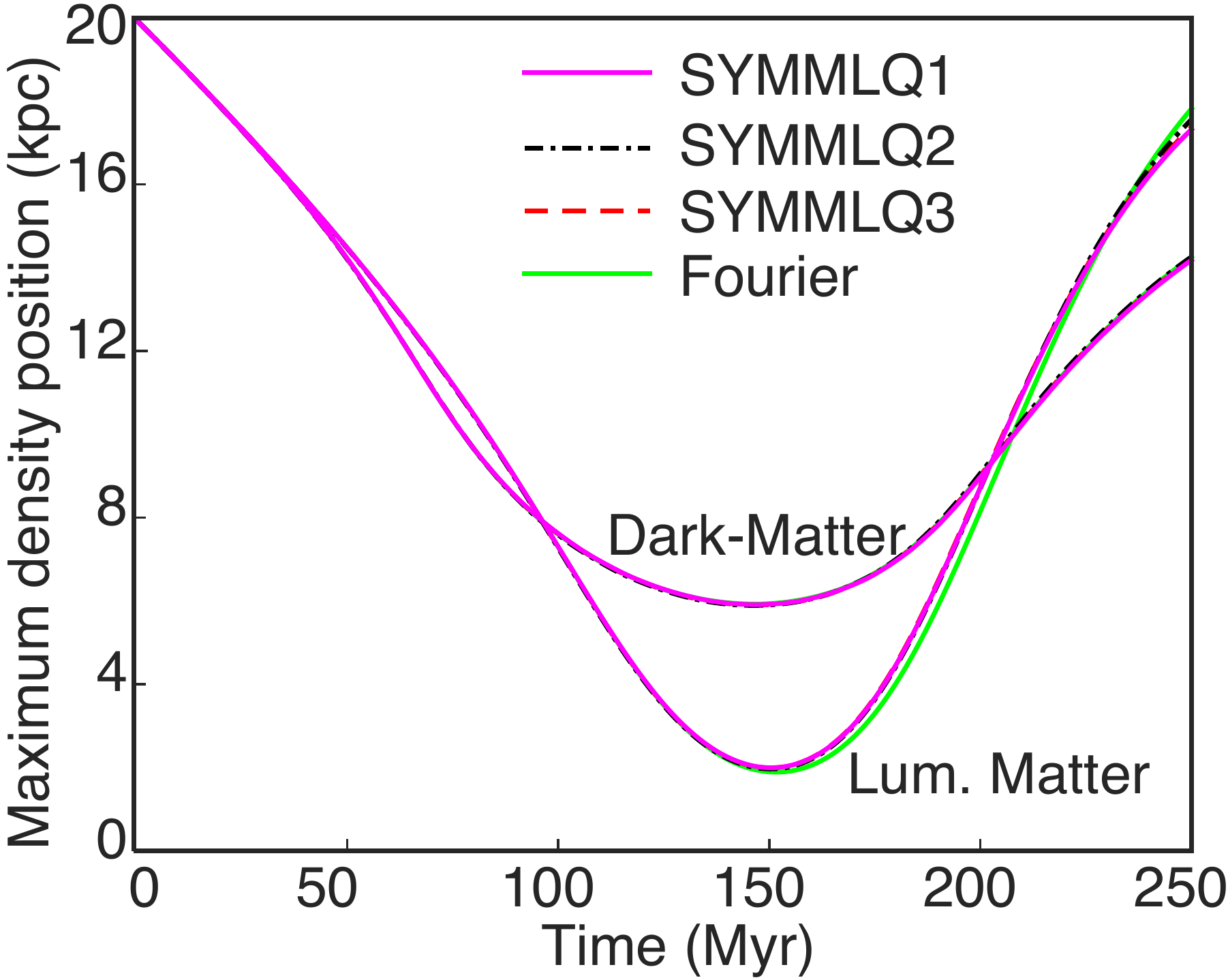}
\includegraphics[width=.49\columnwidth]{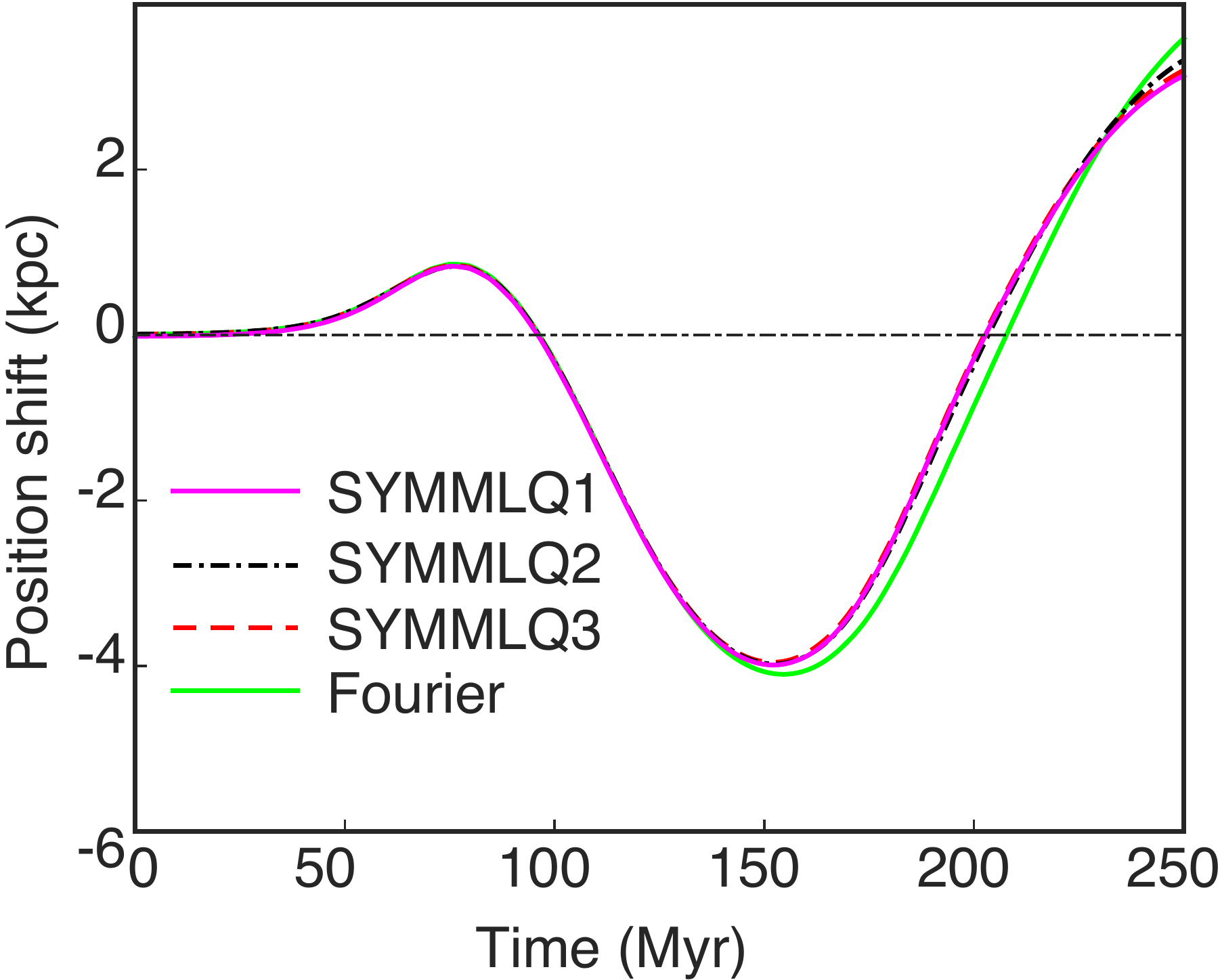}
\end{center}
\caption{\label{suppl_fig1}
Left: comparison of the trajectories of the centers of dark-matter solitons and point-like galaxies obtained with four 
different numerical methods for Eq. (\ref{SPadim}). 
Symmlq1, symmlq2 and symmlq3 use a finite difference scheme to integrate Poisson equation
at each step, with different spatial and temporal grids. For the Fourier line, a Fourier 
transform is used to solve Poisson equation.
Right: Comparison of the offset calculated with the four 
different numerical methods
}
\end{figure}

In order to further check the accuracy and validity of the 
computational methods, we have performed 
a series of auxiliary numerical simulations. In particular, 
we have calculated the effect on the dark matter soliton motion of 
 different numerical schemes for integrating Eq. (\ref{SPadim}).  Results
 for the example corresponding to figure 1 ($\Delta \phi=\pi$)
  are shown in fig. \ref{suppl_fig1},
where we plot the evolution of the DM soliton clouds and the ordinary
matter and the corresponding offset. 
The maximum density position of the dark matter distribution is found by quadratic interpolation
around the maximum value in the discrete grid.
As it can be seen in the pictures, the differences between the methods are negligible for the evolution times
that we have used in the paper.

\subsection* {A fluid toy model for ordinary matter}

\begin{figure}[h!]
\begin{center}
\includegraphics[width=.5\columnwidth]{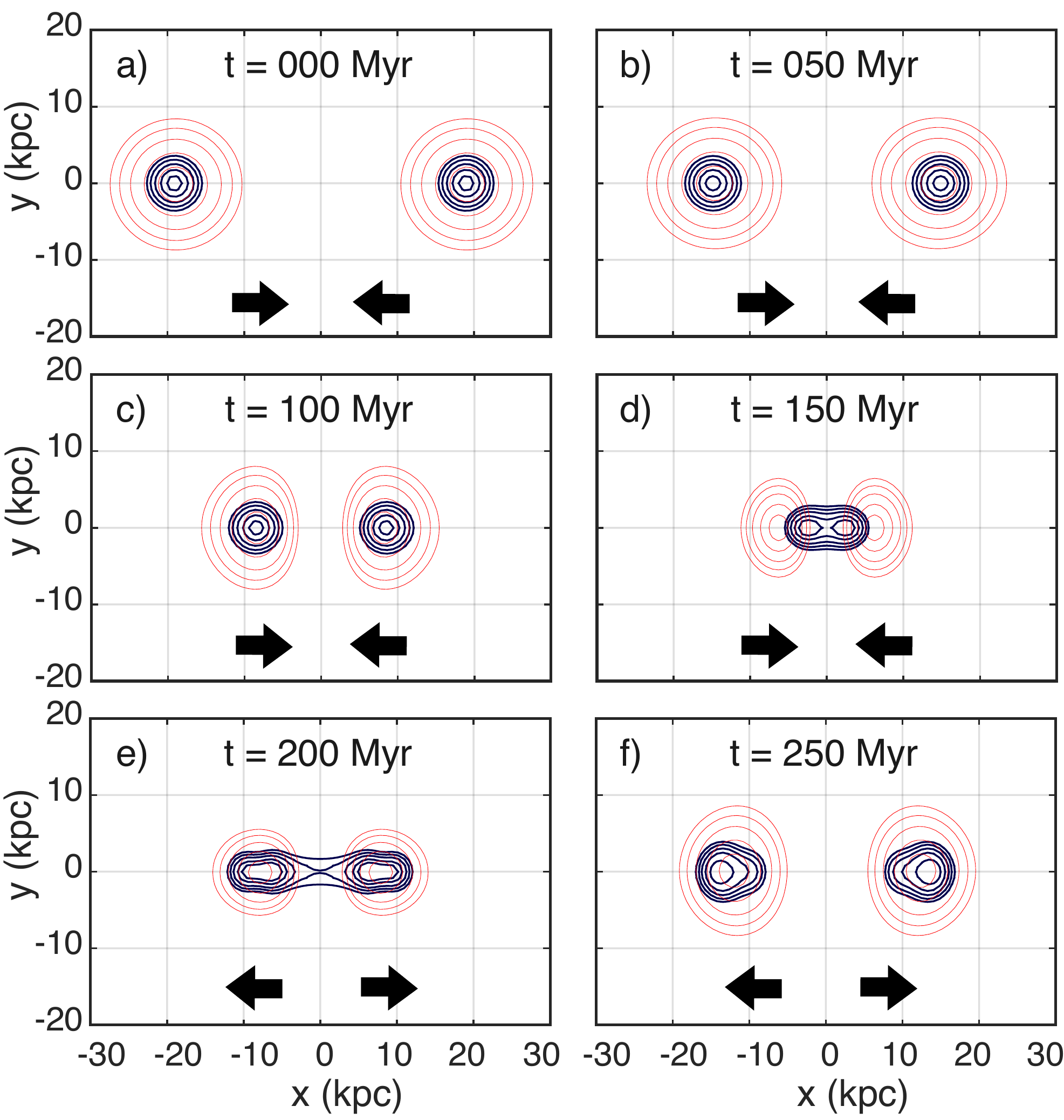}
\end{center}
\caption{Simulation of the evolution of a widespread distribution 
of luminous matter (in black), showing a qualitative 
behavior similar to the  point 
particle model. For the simulation, we used a system of two 
mutually incoherent nonlinear Schrš\"odinger equations, coupled to the system 
of Eq (\ref{SPadim})
 given in the paper. The trajectory of the center of 
 gravity of the luminous matter has been tracked and compared with the evolution of
the "point galaxies" in the inset of fig. 2, with good qualitative agreement.
Red lines correspond to dark matter projected density contours.}
\label{suppl_fig4}
\end{figure}

In this paper, we have used the simplest possible toy model for the stars, considering
just test particles. More realistic descriptions would take into account that luminous matter
is not point-like and that it sources the gravitational potential.
However, these effects should not affect the qualitative conclusions presented above, 
since the offsets are generated by the interference force acting on the dark matter
solitonic cores while not affecting the ordinary matter.
As a first test of this assertion, we have repeated the simulation for the collision in
phase opposition (figs. 1 and 2) by considering the, arguably, simplest
fluid model for ordinary matter: each galaxy is modeled by an independent 
Schr\"odinger equation coupled to Eq. (\ref{SPadim}). Thus,
$i\,\partial_t g_i= - \frac{1}{2\gamma }\nabla^2 g_i + \gamma \Phi g_i $,
where $g_i$ correspond to the luminous matter distribution of each galaxy and
$\gamma$ is a parameter controlling its size ($\gamma=2$ was taken for the figure).
The gravitational potential $\Phi$ is still determined by Eq. (\ref{SPadim}), being
the ordinary matter distributions $g_i$ considered as test fields.
It is well known that 
nonlinear Schr\"odinger equations can be recast as fluid equations using a Madelung
transformation (see, {\it e.g.} \cite{fluid_nlse}). Results are displayed in figure
\ref{suppl_fig4}.

\section*{Acknowledgements} 

We thank D. Olivieri, J. Redondo and J. R. Salgueiro for 
useful comments.
This work is supported by grants FIS2014-58117-P and FIS2014-61984-EXP
from Ministerio de Ciencia e Innovaci\'on. The work of 
A.P. is also supported by the Ram\'on y Cajal program and grant 
EM2013/002 from Xunta de Galicia.

\end{document}